\documentclass[aps,prb,amsmath,amssymb,floatfix,12pt]{revtex4-1}
\usepackage{tabularx}
\usepackage{bm}
\usepackage{euscript}
\usepackage{graphicx}
\usepackage{color}
\usepackage{amsfonts}
\usepackage{exscale}
\usepackage{amsbsy}
\usepackage{subfigure}
\usepackage{textcomp}

\numberwithin{equation}{section}

\newcommand{\sgn}{\mathrm{sgn}}
\newcommand{\nn}{\nonumber}

\newcommand{\be}{\begin{eqnarray}}
\newcommand{\ee}{\end{eqnarray}}
\newcommand{\tr}{\textrm{tr}}
\newcommand{\CL}{{\cal L}}
\newcommand{\CI}{{\cal I}}
\newcommand{\CO}{{\cal O}}

\begin{document}

\title{Non-Fermi liquid behavior of large $N_B$ quantum critical metals}
\author{A. Liam Fitzpatrick$^{{\bar \psi},\psi}$, Shamit Kachru$^{{\bar \psi},\psi}$, Jared Kaplan$^{\phi}$, S. Raghu$^{{\bar \psi},\psi}$}
\affiliation{$^{\bar \psi}$Stanford Institute for Theoretical Physics, Stanford University, Stanford, California 94305, USA}
\affiliation{$^\psi$SLAC National Accelerator Laboratory, 2575 Sand Hill Road, Menlo Park, CA 94025}
\affiliation{$^\phi$Department of Physics and Astronomy, Johns Hopkins University, Baltimore, MD 21218}

\date{\today}

\begin{abstract}

 The problem of continuous quantum phase transitions in metals involves critical bosons coupled to a Fermi surface.  We solve the theory in the limit of a large number, $N_B$, of bosonic flavors, where the bosons transform in the adjoint representation, while the fermions
are in the fundamental representation of a global $SU(N_B)$ flavor symmetry group.  The  leading large $N_B$ solution  corresponds to a non-Fermi liquid coupled to Wilson-Fisher bosons.  
In a certain energy range, 
the fermion velocity vanishes - resulting in the destruction of the Fermi surface.  Subleading $1/N_B$ corrections correspond to a  qualitatively different form of Landau damping of the bosonic critical fluctuations.  We discuss the model in $d=3-\epsilon$ but because of the additional control afforded by large $N_B$, our results are valid down to $d=2$.  In the limit $\epsilon \ll 1$, the large $N_B$ solution is consistent with the  RG analysis of Ref.\onlinecite{Fitzpatrick2013}.
\end{abstract}

\maketitle

\section{Introduction}

 The problem of understanding metals in the vicinity of a quantum phase transition is of significant interest\cite{Stewart2001,Belitz2005,Lohneysen2007,Sachdev}.  Near the transition, gapless bosons (representing the order parameter fields associated with the phase transition) interact with the Fermi surface of the itinerant electrons.  Similar phenomena occur when metals are coupled to gapless gauge bosons: example realizations include gapless spin liquids, and electrons in half-filled Landau levels.  While significant progress has been made in this field\cite{Hertz1976,Holstein,Millis1993,Polchinski1994,Nayak1994,Nayak1994a,Altshuler1994,Chakravarty1995,Oganesyan2001,Abanov2004, Rech2006,Lee2009,Senthil2009,Mross2010,Metlitski2010,Lee2013,Phillips2013}, key aspects of the field theory of critical bosons interacting with a Fermi surface remain poorly understood.  

By considering a generalization of the theory to a large number $N_B$ or $N_F$ of boson or fermion flavors, respectively, we gain additional theoretical control and new insights into the problem in a non-perturbative context.  In such limits, particles with the large number of flavors acts as a dissipative bath for the remaining degrees of freedom.  For example, in the limit of large $N_F$, with $N_B=1$, fermion fields act as the dissipative bath for the order parameter fields; at leading order, the damping of the boson is the most important effect.  In this regime, the IR behavior is closely related to the RPA theory of Hertz and related subsequent work.   On the other hand, when $N_B$ is large and $N_F = 1$, the large number of boson flavors strongly dress the fermions, resulting in the destruction of the Landau quasiparticle at leading order.  Thus, the two limits represent extremes where the IR behavior appears to be qualitatively different.  While much work has been done in the large $N_F$ limit, the large $N_B$ limit remains largely unexplored.  In this paper, we solve the theory at leading order in large $N_B$. 

In \onlinecite{Fitzpatrick2013}, the authors took a Wilsonian approach to this theory, studying the RG flow of the couplings in a controlled perturbative regime at high energies.  
Working in $3-\epsilon$ dimensions, the one-loop analysis yields a theory of a non-Fermi liquid (with anomalous dimensions
that vanish as $\epsilon \to 0$) interacting with a scalar at the Wilson-Fisher fixed point.  The 
analysis of \onlinecite{Fitzpatrick2013} was argued to be valid only at energies intermediate between a low-energy scale $E_{\rm breakdown}$ and the $E_{\rm Fermi}$ scale, where $E_{\rm breakdown}$ is suppressed by a power of $\epsilon$.    

The physics setting the scale $E_{\rm breakdown}$ in our philosophy is the Landau damping of the bosons by the
fermion, due to loop diagrams such as Figure \ref{fig:FermionLoop}.  In perturbation theory about a free UV fixed point, this generates a contribution to the boson self-energy $\Pi(q_0,q) \sim
g^2 k_F^{d-1} ({q_0 \over q}) \theta(q - q_0)$; the coefficient is easily understood as a loop factor multiplying the density of states
at the Fermi surface into which the boson may decay.  This contribution to $\Pi$ cannot appear in a Wilsonian effective action,
as it is non-local, and there are no RG counterterms associated with it.  This is confirmed by an explicit computation, which shows that diagrams such as Figure \ref{fig:FermionLoop} are UV finite.  As a result, Landau damping effects are not well captured by RG.  When damping becomes important, competing with the tree-level terms in the boson propagator, our perturbation theory is rendered invalid.
In the epsilon expansion, this happens at energy scales suppressed by a power of $\epsilon$ as compared to $E_{\rm Fermi}$.
In order to further suppress this energy scale, as well as to avoid various subtleties associated with four-Fermi interactions, a large $N_B$ limit (with bosons in the adjoint of a flavor group and fermions in the fundamental) also proves useful.  

At large $N_B$, one can use $1/N_B$ as the small parameter instead of $\epsilon$.  It is natural to ask if one can solve these models at large $N_B$ but with $\epsilon=1$, i.e. in $d=2$.  In this paper we will solve for the physics of the non-Fermi liquid in $d=2$ at large $N_B$.  We find that under RG evolution, the fermion velocity $v$ decreases in the IR, leading to interesting new physics. 

\begin{figure}
\label{ShamitFigure}
\begin{center}
\includegraphics[width=0.35\textwidth]{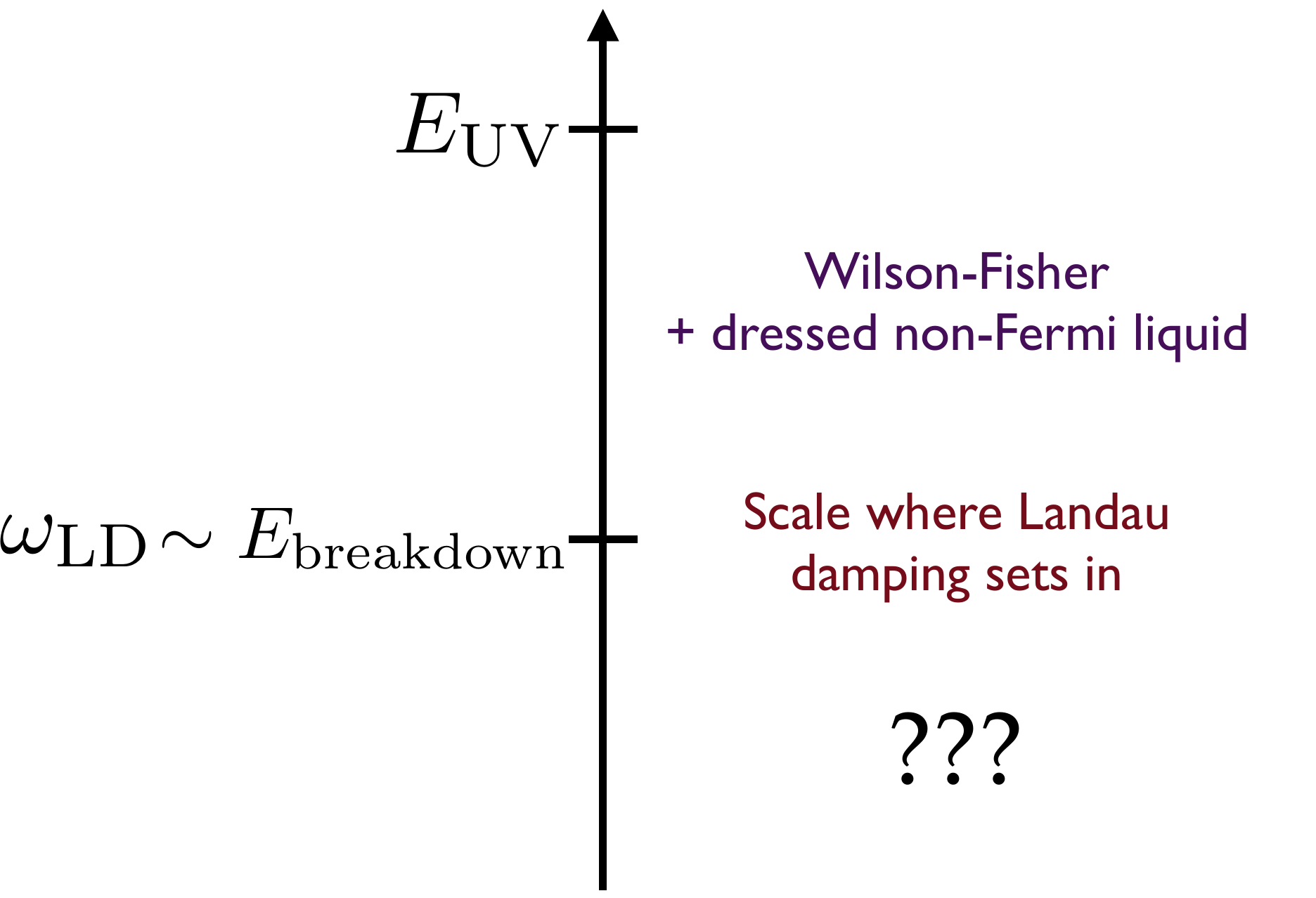}
\end{center}
\caption{The fixed points we are studying govern physics over a finite range of energies at finite $\epsilon$ and $N_B$, though
in the strict large $N_B$ limit the region of control extends down to arbitrarily low energy.  The scale at which our control breaks down
is the scale associated with significant Landau damping of the bosons.
}
\label{fig:FermionLoop2}
\end{figure}

Our philosophy here and in \onlinecite{Fitzpatrick2013} differs somewhat from that employed in a number of other recent works on this subject.  It would be very interesting to determine the $\omega \to 0$ behavior of a potential non-Fermi liquid emerging from scalar/fermion interactions, with the $\omega \to 0$ limit taken ${\it before}$ any control parameters are taken to extreme values.  This is the goal of many recent works, which aim to find a self-consistent ansatz for such a putative fixed point.  However, this was not our aim in \onlinecite{Fitzpatrick2013}, nor in the present work.  Instead, we are content to use perturbation theory about a controlled UV fixed point (a decoupled Fermi liquid and a self-interacting scalar), and to find approximate fixed points visible in perturbation theory, valid only over a range of scales ${\it not}$ necessarily extending to $\omega \to 0$.  We feel this is of interest for two reasons: firstly, such studies can be performed reliably using standard techniques in field theory for a range of models (varying, for instance, the ratio of the number of bosonic to fermionic degrees of freedom); the differences in the results for different models could be instructive.  Secondly,
in many cases, the $\omega \to 0$ limit is anyway inaccessible, as new phases intervene in the attempt to study the ultra low-energy behavior.  As a case in point, superconducting instabilities are expected to intervene in many models of the type we discuss.

An outline of this paper is as follows.
First, in \S2, we review the RG structure described in \onlinecite{Fitzpatrick2013}, and discuss a subtle issue of
scheme dependence.  We also summarize the physics of the approximate fixed point at small $\epsilon$.
In \S3,  we derive a large $N_B$ solution using the ``gap equation,'' a self-consistent integral equation for the fermion self-energy
$\Sigma$ which incorporates the physics of all of the ``rainbow diagrams'' (see Figure \ref{fig:GapEquation}).
Next, in \S4, we again solve the theory at leading order in large $N_B$ by a different technique, using a trick to recursively evaluate the perturbative corrections to $\Sigma$ to all orders.
The agreement between
the methods of \S3\ and \S4, and the agreement of both with the $\epsilon$ expansion results
of \S2 and \onlinecite{Fitzpatrick2013}, give us confidence in the consistency of the large $N_B$ solution.
In \S5, we turn to the leading large $N_B$ correction of greatest interest -- the boson self-energy diagram in Figure 2, which contributes to Landau damping of the bosons.  We find that the form of the Landau damping due to the non-Fermi liquid is qualitatively different from the damping imparted by a conventional Fermi liquid.  This gives further justification for our procedure of
studying intermediate fixed points starting from the UV action, because the emergent IR physics of Landau damping depends crucially
on the modifications to the Fermi liquid that occur at intermediate scales.  Finally, in \S6, we provide a more detailed discussion of the subtleties
associated with
regulator choices.  This section can be read independently of \S3-5.  \S7 contains our conclusions and a discussion of the
larger picture we see emerging from these kinds of studies.

\begin{figure}
\begin{center}
\includegraphics[width=0.35\textwidth]{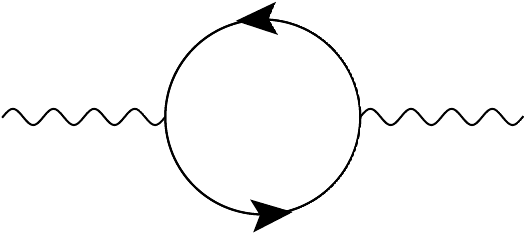}
\end{center}
\caption{ This figure depicts the $1/N_B$ suppressed one-loop fermion diagram that renormalizes the boson propagator. It differs greatly between a weakly coupled fermi-liquid and our approximate fixed-point theory in $d=2$ dimensions at large $N_B$, leading to rather different conclusions concerning the effect of Landau damping. }
\label{fig:FermionLoop}
\end{figure}

\section{RG Structure in an expansion in $\epsilon=3-d$ and $1/N_B$}

\begin{figure}
\begin{center}
\includegraphics[width=0.48\textwidth]{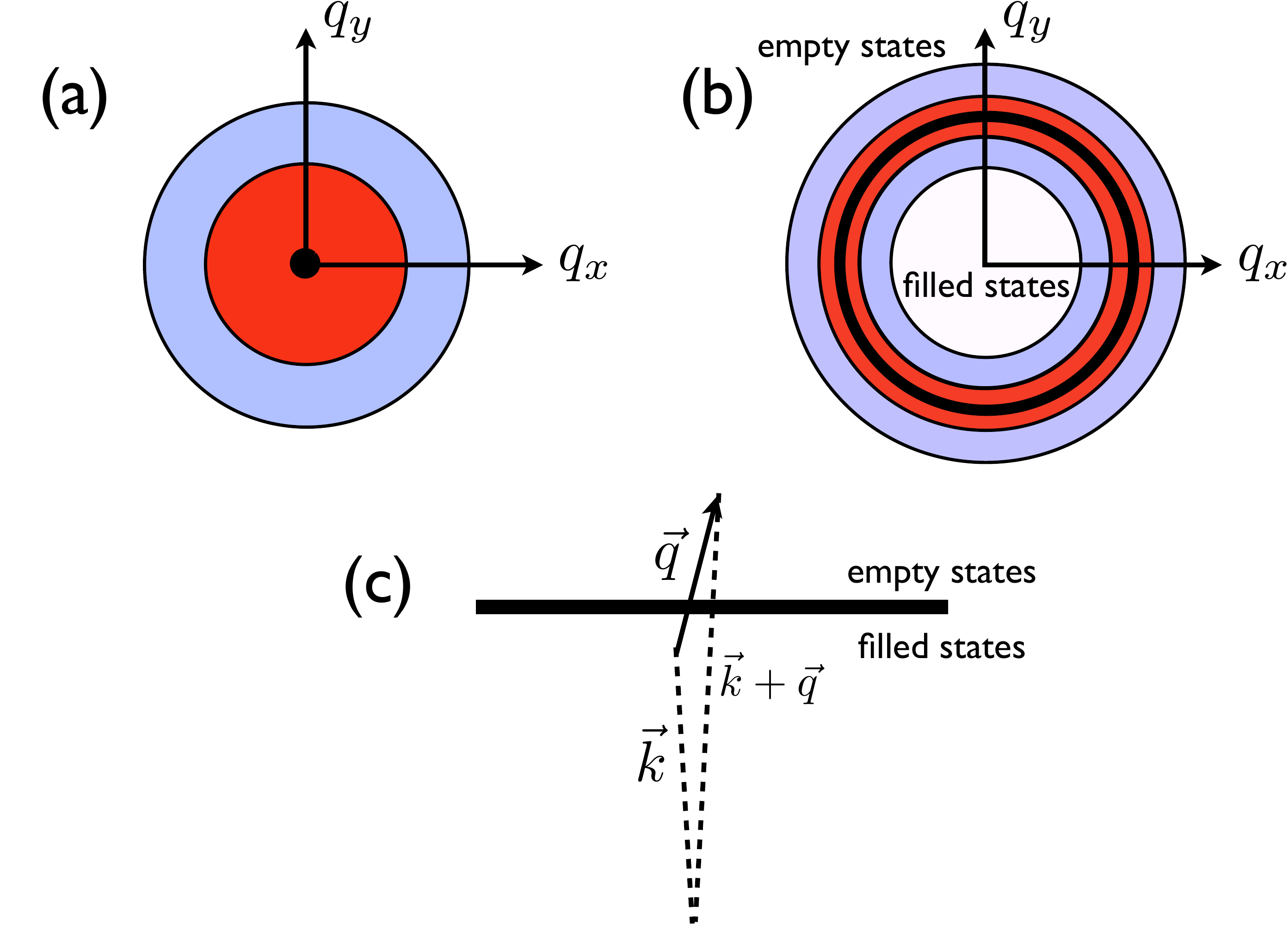}
\end{center}
\caption{Summary of tree-level scaling.  High energy modes (blue) are integrated out at tree level and remaining low energy modes (red) are rescaled so as to preserve the boson and fermion kinetic terms.  The boson modes (a) have the low energy locus at a point whereas the fermion modes (b) have their low energy locus on the Fermi surface.  The most relevant Yukawa coupling (c) connects particle-hole states  separated by small momenta near the Fermi surface; all other couplings are irrelevant under the scaling.   
}
\label{fig:scaling}
\end{figure}

\begin{figure}
\begin{center}
\includegraphics[width=0.45\textwidth]{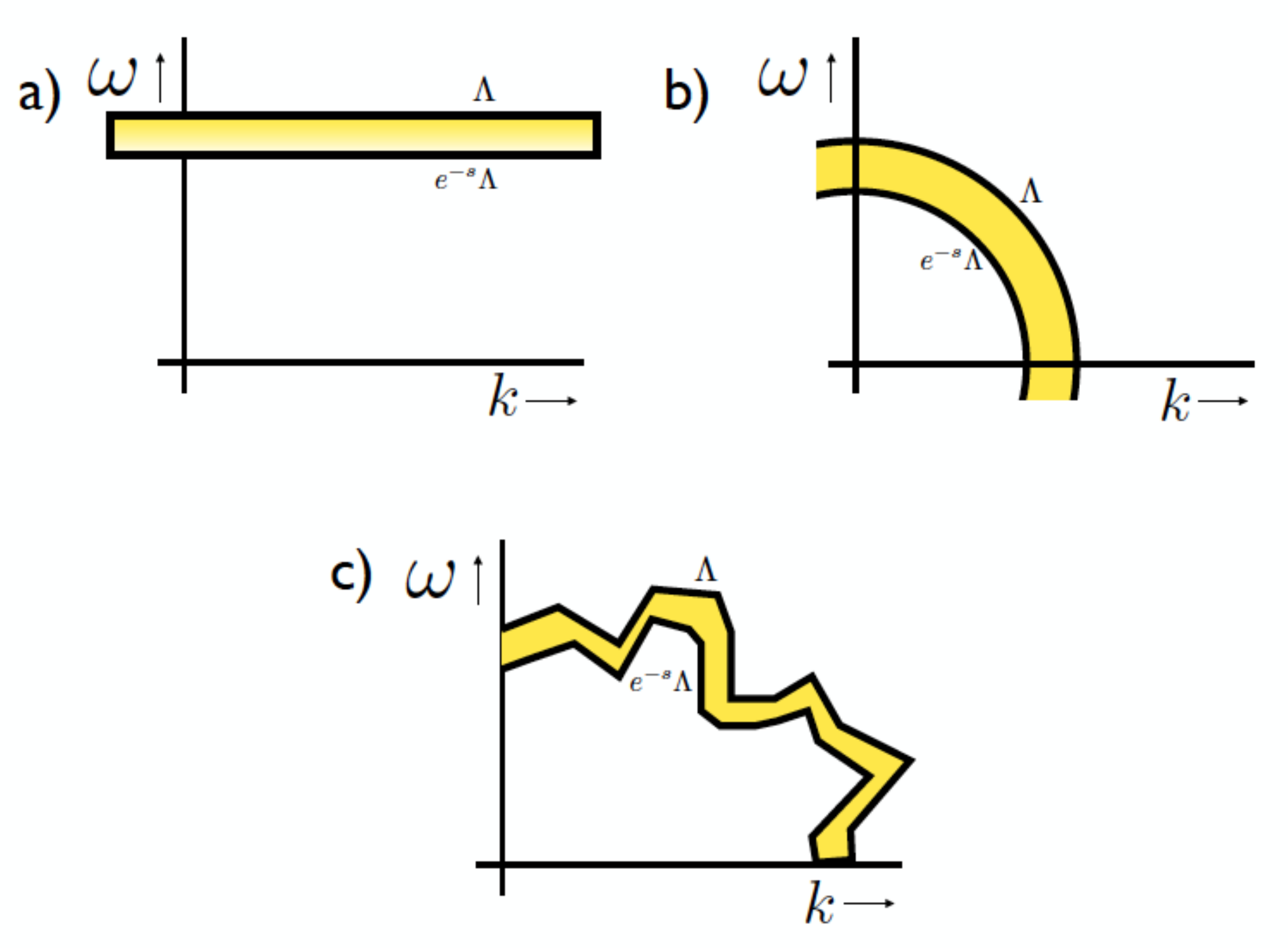}
\end{center}
\caption{Examples of possible schemes for decimating high energy modes.  Scheme a), which we
adopted in \onlinecite{Fitzpatrick2013}, integrates out shells in $\omega$ but integrates out all momenta
in a given frequency shell.  Scheme b), which integrates out both frequencies and momenta, is better for our purposes (as explained below and in \S6), and we adopt it in this paper.  Scheme c) is recommended as an assignment for graduate students one wishes to avoid.}
\label{fig:RGschemes}
\end{figure}

Our focus is on the field theory with Lagrangian

\begin{eqnarray}
\mathcal L_{\psi} &=&   \bar \psi^i \left[ \partial_{\tau}  + \mu -\epsilon(i  \nabla)  \right] \psi_{i} 
+ \frac{\lambda_{\psi} }{N_B}\bar \psi^i \psi_{i} \bar \psi^j \psi_{j}
\nonumber  \\
\mathcal L_{\phi} &=& {\rm tr}\left( m_{\phi}^2   \phi^2 + \left(\partial_{\tau} \phi \right)^2 + c^2 \left( \vec \nabla \phi \right)^2\right)   \nonumber \\
&&+  \frac{\lambda_\phi^{(1)}}{8 N_B} \tr (\phi^4) + \frac{\lambda_\phi^{(2)}}{8 N_B^2} (\tr (\phi^2))^2  \nonumber \\
\mathcal L_{\psi,\phi}&=& \frac{g}{\sqrt{N_B}}   \bar \psi^i \psi_{j} \phi^j_i
\label{largeNaction}
\end{eqnarray}
The (spinless) fermions are in an $N_B$-vector $\psi_i$, while the scalar $\phi^j_i$ is an 
$N_B \times N_B$ complex matrix.  We take the global symmetry group to be $SU(N_B)$, and as in
\onlinecite{Fitzpatrick2013}, we will set $\lambda_{\phi}^{(1)} = 0$.  This choice is technically natural; if $\lambda_\phi^{(1)}$ is set to zero in the UV, then it is never generated by radiative corrections 
(this can be understood simply as a consequence of an enhanced $SO(N_B^2)$ symmetry at $\lambda_{\phi}^{(1)}=0$ broken softly by the Yukawa coupling).
Furthermore, it makes the analysis far more tractable.  

In this section, we describe the perturbative RG approach to studying this system, following \onlinecite{Fitzpatrick2013}.
We start with the same RG scaling as in that paper, scaling boson and fermion momenta differently as in Figure 3 (in a way
that is completely determined by the scaling appropriate to the relevant decoupled fixed points at $g=0$).
However, we depart in one important way from the philosophy of \onlinecite{Fitzpatrick2013} - instead of decimating
in $\omega$ between $\Lambda$ and $\Lambda - d\Lambda$ at each RG step, but integrating out all momenta (as in
Figure 4 a)),
we instead do a more `radial' decimation, integrating out shells in both $\omega$ and $k$ (as in Figure 4 b)).  This
introduces two UV cutoffs in the problem, $\Lambda$ and $\Lambda_k$.  We find this procedure superior because it
avoids the danger of retaining very high momentum modes %(at large $k$) 
at late steps of the RG.  
While of course observables
will agree in the different schemes, aspects of the physics which are obscure in the scheme of Figure 4 a) become manifest
in the scheme we have chosen here.  This elementary (but sometimes confusing) point is discussed in more detail in \S6, which can
be read more or less independently of the rest of the paper.

The large $N_B$ RG equations are quite simple.  The Yukawa vertex renormalization is a ${\cal O}(1/N_B)$ effect, and the boson wave-function
renormalization due to fermion loops is UV finite and thus does not contribute to the RG equations.
Therefore, at leading order, the boson is governed
by an $O(N_B^2)$ Wilson-Fisher fixed point, while the fermion wave-function renormalization governs the non-trivial beta
functions.  Here and throughout the paper we will use the notation `$\ell$' to represent the component of the fermion momentum perpendicular to the fermi surface and `$\omega$' to represent fermion energies.    Writing
\be
\CL_\phi &=& \phi^2(\omega^2 + c^2 k^2), \nn\\
\CL_\psi &=& (1+ \delta Z)  \psi^\dagger i \omega \psi - (v+ \delta v) \psi^\dagger \ell \psi , \nn\\
\CL_{\psi\phi} &=&  (g+ \delta g) \phi \psi^\dagger \psi , \nn\\
\delta Z &\equiv& Z-1, \qquad \delta v \equiv v_0 Z - v, \qquad \delta g\equiv g_0 Z -g~,
\ee
we simply need to compute the logarithmic divergences in $\delta Z$ and $\delta v$ to find the one-loop running.  
$\delta Z$ and $\delta v$ are chosen to cancel the log divergences in the one-loop self-energy $\Sigma$.  Performing
the explicit computation, we find
\be
\Sigma &=& a_k \log (\Lambda_k) + a_\Lambda \log (\Lambda )+ a_E \log (E), \nn\\
a_\Lambda &=& -\frac{b g^2}{c |v| (c+|v|)} (i \omega - v \ell), \nn\\
a_E &=& -\frac{b g^2}{c^2 (c+|v|)} (i \omega +\sgn(v) c \ell) , \nn\\
a_k &=& \frac{b g^2}{c^2|v| } i  \omega ~.
\label{eq:CartOneLoop}
\ee 
where $b$ is an ${\cal O}(1)$ (positive) number, computed in appendix \ref{app:OneLoop}. One can verify that at $v$=0, $\sgn(v)=0$ in the above formula by taking $v=0$ inside the loop integral. Next, one chooses counter-terms
to cancel the dependence on the cut-off $\Lambda$; this is equivalent to setting a UV boundary condition for the parameters of the theory.  The dependence on $\Lambda$ is eliminated if we take the following counter-terms:
\be
\delta Z &=& \frac{b g^2}{c |v|}  \left( \frac{1}{c+|v|} \log \Lambda-\frac{1}{c} \log \Lambda_k\right) \nn\\
\delta v &=&  \frac{b g^2}{c (c+|v|)} \sgn(v)\log \Lambda \nn\\
\delta g&=&0, \nn\\
g_0 &\approx& g +\delta g- \delta Z g, \qquad v_0 \approx v +\delta v- \delta Z v.
\ee
As in \onlinecite{Fitzpatrick2013}, the four-Fermi terms have a stable fixed point at $\lambda_{\psi}=0$ and we do not
discuss them further here.

We will define beta functions by setting $\Lambda_k \sim \Lambda$ and computing running with respect to $\Lambda$.
The results for the beta functions and the anomalous dimension of the fermion are:
\be
\beta_g &\equiv&  {\partial g \over \partial {{\rm log} \Lambda}}= g \left( -\frac{\epsilon}{2} + \frac{b g^2}{c^2 (c+|v|)} \right) ,
\label{eq:OneLoopBeta1}\\
\beta_v &\equiv& {\partial v \over \partial {{\rm log}\Lambda}}=  \frac{b g^2}{c^2} \sgn(v), \label{eq:OneLoopBeta2} \\
2\gamma &\equiv - {\partial \delta Z \over {\partial {\rm log} \Lambda}}&=  \frac{b g^2}{c^2 (c+|v|)}. ~\label{eq:OneLoopBeta3}
\ee

As a check that $b>0$, note that the anomalous dimension is positive, as is guaranteed by unitarity at the fixed point.

Several pieces of important physics are evident in (\ref{eq:OneLoopBeta1}) - (\ref{eq:OneLoopBeta3}):

\medskip
\noindent
$\bullet$ There is a controlled fixed point at $g$ of order $\sqrt{\epsilon}$, where the fermions are dressed into a
non-Fermi liquid.

\medskip
\noindent
$\bullet$ The anomalous dimension of the fermion is ${\epsilon \over 4}$, in agreement with the result in \onlinecite{Fitzpatrick2013}.\footnote{Note that the anomalous dimension is larger by a factor of 2 at large $N_B$ than at $N_B=1$.  This is because the vertex correction is non-planar and does not contribute at large $N_B$.}
The Green's function for the fermion satisfies a Callan-Symanzik equation
\be
 \left( \Lambda \frac{\partial}{\partial \Lambda} + \beta_g \frac{\partial }{\partial g} + \beta_v \frac{\partial }{\partial v} + 2 \gamma \right) G_F\left( \frac{\omega}{\Lambda},\frac{\omega}{ \ell}; g, v\right)  ~=~0~.
\ee
Therefore, at the fixed point where the beta functions vanish, it will take the form
\begin{equation}
G_F ~={~{1 \over \omega^{1- {\epsilon \over 2}}}} ~f({\omega \over \ell})~.
\end{equation}
We will see in the next sections that the scaling function $f$ in the large $N_B$ theory is trivial -- $f \sim {\rm const}$.
This is completely consistent with the result in \onlinecite{Fitzpatrick2013}, but indicates that in the RG scheme used there,
the scaling function (which was not determined in that paper) is non-trivial.

\medskip
\noindent
$\bullet$ We save the most interesting point for last.  The beta function for the Fermi velocity is such that the velocity flows to
${\it zero}$ in the IR.   
A calculation of (\ref{eq:OneLoopBeta2}) in the $v=0$ theory confirms that the beta function for $v$ vanishes there.  The physics of vanishing $v$ should be interpreted
cautiously.  In the full high-energy theory, there are further corrections to the action involving higher spatial derivatives.  For
instance, one would have terms of the form
\begin{equation}
\delta {\cal L} \sim {\ell^2 \over 2m^*} \psi^\dagger \psi~,
\end{equation}
with $m^*$ a UV mass scale related to the curvature terms at the Fermi surface.  While these are ${\it irrelevant}$ operators in the theory with finite Fermi velocity, as $v$ flows
to be very small, the role of such terms in the fermion propagator becomes more important.  There is a cross-over from a
theory of a normal Fermi surface to a `$z=2$ scaling' governed by the quadratic term at very low energies where
$v \to 0$.  The physics is characterized by fermions which have become so `heavy' that they can no longer be created (even
in virtual pairs in loops), and therefore are similar to `non-relativistic' fermions at zero density. Consequently, for $v$ exactly zero, diagrams with fermion loops vanish.  However, because the transition to $z=2$ scaling occurs only when $v \ell$ is less than $\frac{\ell^2}{2m_*}$, it is important whether $v$ runs exactly to zero or only to a value proportional to $\frac{1}{m_*}$, which has so far been set to zero in our analysis.  In appendix \ref{app:Loops}, we analyze the running at $v \ll \frac{\Lambda}{m_*}$, and argue that at finite $m_*$, $v$ does not run to be smaller than $\CO(\frac{\Lambda}{m_*})$, where $\Lambda$ is the cut-off.  In appendix \ref{app:LDsmallv}, we compute Landau damping as an example of a fermion loop that is suppressed at small $v$, and show that at small $v$ the small dimensionless parameter suppressing the fermion loop is $\frac{v m_*}{q}$, where $q$ is the external boson momentum.  Since $v$ runs to be $\CO(\Lambda/m_*)$ at large but finite $m_*$, it does not become small enough to suppress fermion loops in practice in a Wilsonian treatment.

In the rest of the paper, we will formally solve the $v \to 0$ fixed point.  This is the formally correct thing to do in the limit
that the high mass scale $m^* \to \infty$.  The interesting physics of the cross-over to the theory with $z=2$ scaling 
when $m_*$ is large but finite is left for further exploration.

\section{Large $N_B$ Solution from Gap Equation}

At large $N_B$, the fermions have small backreaction on the boson, so the leading large $N_B$ behavior of the boson Green's function is governed by the $O(N_B^2)$ Wilson-Fisher fixed point:
\begin{equation}
\langle \phi(p) \phi(-p)\rangle = {1\over p^{2(1-\gamma_{\phi})}} + \mathcal O(1/N_B)
\end{equation}
with
\begin{equation}
\gamma_\phi = {\epsilon^2 \over 4N_B^2}~.
\end{equation}
This is a result of the simplification in our action afforded by setting $\lambda_{\phi}^{(1)} = 0$; otherwise the leading
large $N_B$ bosonic theory would also be highly non-trivial.

\begin{figure}
\label{rainbows}
\begin{center}
\includegraphics[width=0.35\textwidth]{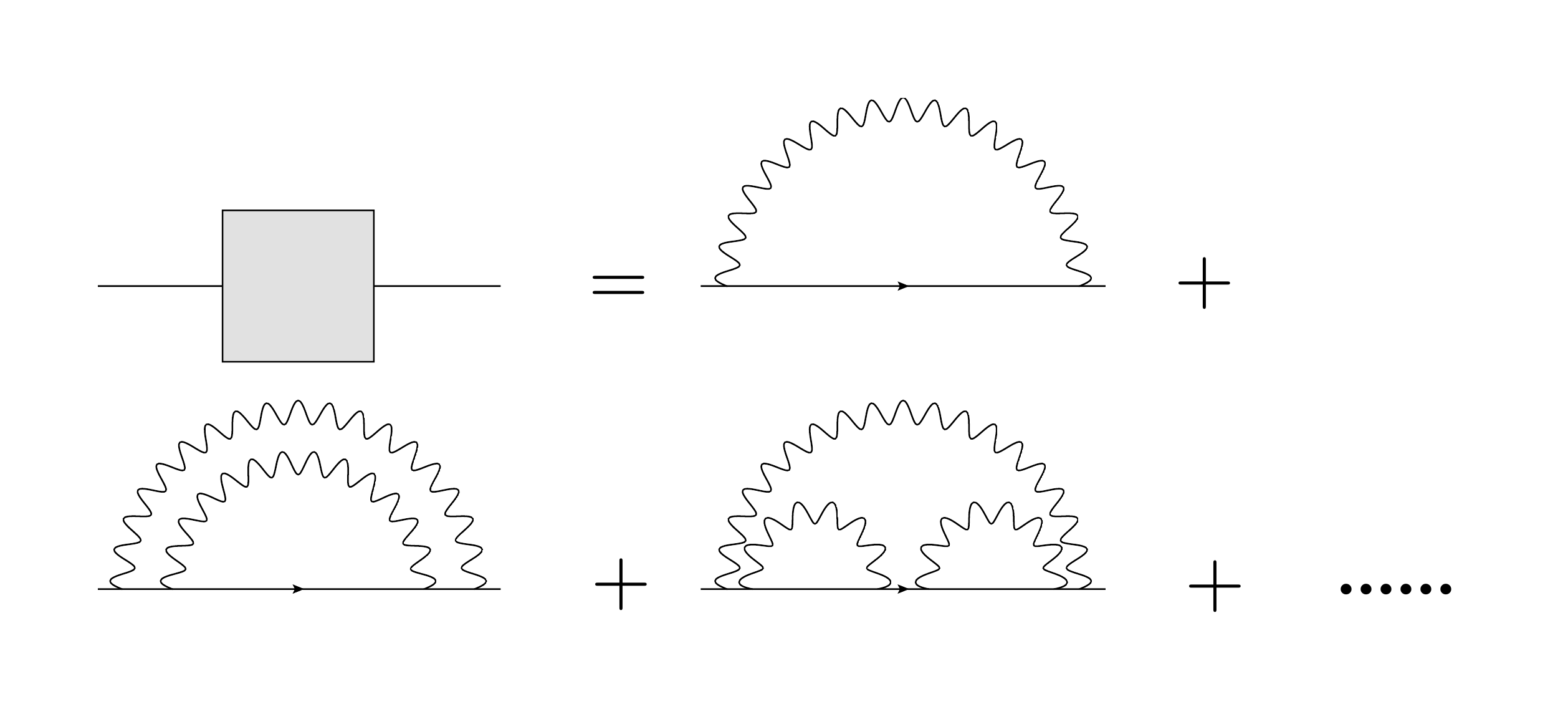}
\end{center}
\caption{The rainbow diagrams which determine the fermion self-energy at large $N_B$.}
\label{fig:GapEquation}
\end{figure}

However, for the fermion, the full set of rainbow diagrams (\ref{rainbows}), depicted in Figure \ref{fig:GapEquation}, contribute to the self-energy.  As is standard
in large $N_B$ theories (see e.g. \onlinecite{Coleman}), this results in a gap equation for the fermion self-energy
\be
\label{gapeqn}
\Sigma(\omega_e, \ell_e) &=&g^2 \int \frac{d\omega d \ell d^{d-1} k_\parallel}{(2\pi)^{d+1}} \frac{1}{(\omega^2 + \ell^2 + k_\parallel^2)(i (\omega - \omega_e) - v_F (\ell - \ell_e) + \Sigma(\omega - \omega_e, \ell - \ell_e))} \nn\\
\ee

The solution of the gap equation at the large $N_B$ fixed point discussed in \S2\ is very simple.  In the idealized limit
where $v_F \to 0$, and neglecting the $\ell^2 / 2m^*$ corrections, the fermion Green's function can only be a function
of $\omega$.  Allowing for a general fermion anomalous dimension $\gamma$, we take the Green's function to be 
\begin{equation}
G_F = {\mu^{-2\gamma} \over \omega^{1-2\gamma}}~,
\end{equation}
where $\mu$ is an RG scale introduced to satisfy dimensional analysis. Therefore, the self-energy is expected to be of the form
\begin{equation}
\Sigma(\omega,k) = - \omega + G^{-1}_F~.
\end{equation}
We can see that at very low external frequencies, the left-hand side of the gap equation (\ref{gapeqn}) is dominated by
$G_F^{-1}(\omega_e)$.  The resulting equation:
\be
\frac{\omega_e^{1-2\gamma}}{\mu^{-2\gamma}} &=& g^2 \int \frac{d \omega d \ell d^{2-\epsilon} k}{(2\pi)^{4-\epsilon}} \frac{1}{\omega^2 + c^2 (\ell^2 + k^2)}\frac{\mu^{-2\gamma}}{(\omega+ \omega_e)^{1-2\gamma}}.
\ee
The scaling of the left- and right-hand sides with respect to $\omega_e$ can be seen by inspection to be consistent if and only if $2\gamma=\frac{\epsilon}{2}$, in agreement with the RG equations (\ref{eq:OneLoopBeta1})-(\ref{eq:OneLoopBeta3}). Setting this value for the anomalous dimension, the above turns into a non-trivial equation for the fixed-point coupling $\overline{g}\equiv g \mu^{-\epsilon/2}$:
\be
\frac{1}{\bar{g}^2} &=& \frac{1}{c^{3-\epsilon}} \int \frac{d \omega d \ell d^{2-\epsilon} k}{(2\pi)^{4-\epsilon}} \frac{1}{\omega^2 + \ell^2 + k^2} \frac{1}{(\omega + 1)^{1-\frac{\epsilon}{2}}}.
\ee
so that one can view $\bar g$ as a function of $c$ and $\epsilon = 3-d$.

\section{Large $N_B$ Solution from Perturbation Theory}

One can also directly solve for the Green's function of the $v_F = 0$ fixed point theory by resumming an iterative perturbation theory.
Using dimensional analysis, and the fact that order by order one finds that the self-energy is independent of $\ell$, one can
expand
\begin{eqnarray}
\Sigma(\omega) &=&  i\omega\sum_{n=1}^\infty b_n(\epsilon) g^{2n} (\omega^2)^{-{\epsilon \over 2}n}\\
&=&i \omega \sum_{n=1}^\infty b_n(\epsilon) Z^n, \ \ \  \mathrm{with} \ \ \ Z\equiv {g^2 \over (\omega^2)^{\epsilon \over 2}}~.
\end{eqnarray}
The Green's function has a similar expansion
\begin{equation}
G_F = - {1\over i \omega} \sum_{m=0}^\infty a_m(\epsilon)Z^m~.
\end{equation}

Plugging into the gap equation (\ref{gapeqn}), we find
\be
 \Sigma(\omega_e) &=& -g^2 \frac{{\rm vol}(S^{d-2})}{(2\pi)^{d+1}} \frac{\pi}{2 \sin(\pi \frac{\epsilon}{2})} \int \frac{d \omega d \ell}{(\omega^2 + \ell^2)^{\frac{\epsilon}{2}} } G(\omega+\omega_e) \nn\\
    &=&  i \omega_e \sum_{m=0}^\infty  Z^m  a_m(\epsilon) B_m(\epsilon) , \label{eq:recursion}
  \ee
where $B_m(\epsilon)$ can be written in closed form in terms of Gamma functions.
At small $\epsilon$, it is approximated by
 \be
  B_m(\epsilon) &=& -\frac{1}{(2\pi)^2 (m+1) \epsilon} + \CO(\epsilon^0).
  \ee
  By matching powers of $Z^m$ in (\ref{eq:recursion}), one obtains a recursion relation for $a_m$.  This equation for $a_m$ can then be solved, yielding
\be
a_m(\epsilon) &=& \left( -\frac{1}{ 2\pi^2 \epsilon} \right)^m \frac{\Gamma(\frac{1}{2}+m)}{\Gamma(\frac{1}{2})m!}.
\ee
Now we can sum the series for the Green's function to find
\be
G(p) 
 &=& -\frac{1}{i \omega} \sqrt{ 1 + \frac{g^2}{2 \pi^2 \epsilon} |\omega|^{-\epsilon}}\nn\\
&\stackrel{\omega \ll g^{2/\epsilon}}{\approx} & 
\omega^{\frac{\epsilon}{2} - 1} \sqrt{\frac{2\pi^2 \epsilon}{g^2}} 
\ee
in perfect agreement with the anomalous dimension computed in \S2\ and the self-consistent solution of the gap equation
guessed in \S3.

\section{$1/N_B$ Corrections and Landau Damping}

Now we would like to study how the fermionic degrees of freedom affect the bosons.  In the perturbative regime with $N_B = 1$, it is well known that fermion loops (as in Figure 2) lead to Landau damping of the bosonic degrees of freedom.  In our theory with $N_B \gg 1$ these effects are suppressed by $1/N_B$, but more importantly, the form of Landau damping changes qualitatively due to the RG evolution to small fermion velocity $v \ll c$ and the finite anomalous dimension of the fermions.

In the limit $v \ll c$ the straight-forward perturbative result for Landau damping acquires a new interpretation, which will be discussed further in \onlinecite{KivelsonFitzpatrick2013}.  This standard result for damping is
\be
\Pi &=& \frac{g^2 k_F}{2 \pi v} \left[ \frac{\omega}{ \sqrt{ \omega^2 - v^2 q^2 } }  - 1 \right]   
\nn \\
& \approx & \frac{g^2 k_F}{2 \pi}  \frac{v \, q^2}{\omega^2},
\ee
where the $-1$ corresponds to a contribution to the boson mass, and it has been included by adding  a local counter-term to the boson action.  When the boson is nearly on-shell, we must have $\omega \approx c q \gg v q$, so we expect that Landau damping simply corresponds to a small local effect in the regime where the theory is under control.  This perspective will be further explored in \onlinecite{KivelsonFitzpatrick2013}. 

The presence of an anomalous dimension in the Green's function for the fermions alters the form of Landau damping.   As $v$ approaches 0, the standard kinetic term $v \ell$ eventually becomes sufficiently small that the higher order term $\frac{\ell^2}{2m_*}$ can be comparable, and the exact Green's function for large but finite $m_*$ depends on a treatment of both terms, which is outside the scope of this paper.  
For illustrative purposes, we will consider two examples for the fermion Green's function motivated by our analysis.  The first is $G^{-1} = (-i \omega + \ell^2/2m_*)^{1-\frac{\epsilon}{2}}$.  
Computing the diagram of Figure \ref{fig:FermionLoop} at large $N_B$ using this fermion 2-pt function, one has
\be
\Pi &=& \frac{g^2}{N_B} \int \frac{ d \omega d\ell k_F^{d-1} d^{d-1} \hat{\Omega}}{(2\pi)^{d+1}} \frac{1}{(-i \omega+ \frac{\ell^2}{2m_*})^{1-\frac{\epsilon}{2}}(-i(\omega + q_0) + \frac{(\ell+ q \cos \theta)^2}{2m_*})^{1-\frac{\epsilon}{2}}} = 0.
\ee
The integral has been performed by first closing the $d \omega$ integration contour in the upper half-plane (this does not change the integral when $d>2$).  The branch cuts of the fermion Green's function are at negative real values of the argument, which one can easily see occur only in the lower half-plane for $\omega$.  The physical interpretation is that for this `$v=0$' choice of the Green's function, the fermion has become very heavy and particle-hole pairs cannot be produced kinematically.  Thus we immediately find that Landau damping vanishes exactly.

Our second illustrative example is to take the fermion Green's function to be $G^{-1} = (-i\omega + v \ell)^{1-\frac{\epsilon}{2}}$, and to take $v \ll c$ at the end of the computation.
We now find for the diagram of Figure \ref{fig:FermionLoop}
\be
\Pi &=& \frac{g^2}{N_B} \int \frac{d \omega d\ell k_F^{d-1} d^{d-1} \hat{\Omega}}{(2 \pi)^{d+1}} \frac{1}{(-i\omega + v \ell )^{\frac{d-1}{2}}} \frac{1}{(-i(\omega + q_0) + v \ell + v \cdot q  )^{\frac{d-1}{2}}}
\nn \\
&=& \frac{g^2 k_F}{2 \pi^2 v N_B}  q_0 \log \frac{q_0}{\Lambda}
\ee
where we have taken $d=2$ and $v \ll c$ in order to compute the last line.  This finite damping will become large compared to the boson kinetic term when the boson energy $q_0 \lesssim \frac{k_F}{N_B}$.  As claimed, the damping takes a parametrically different form than would be obtained in the absence of the anomalous dimension.

\section{RG Scheme Dependence and $k_F$}

Here we will make some general comments about scheme dependence in the non-relativistic RG. This section can be read independently of the rest of the paper, and the reader eager for denouement may unreservedly proceed directly to \S \ref{sec:conclusions}.  The major point is that to perform the RG usefully, we must find a scheme where large logarithms of the form $\log(E_i/k_F)$ do not enter, where $E_i$ is some low-energy momentum scale, e.g. the external frequency $\omega$ or perpendicular momentum $\ell$ in the fermion two-point function.

Correlators typically take the form
\be
{\cal A} &=& E_i^a F\left(\frac{E_i}{k_F}, \frac{E_i}{\Lambda}\right),
\ee
where $a$ is a calculable exponent given by tree-level scaling of the correlator.  Due to logarithmic divergences, the function $F$ will usually have some dependence on the cutoff $\Lambda$ of the form $\log(E_i/\Lambda)$, and thus even at weak coupling, one has large loop corrections in the IR when $E_i/\Lambda$ becomes very small.  Thus, as is well-known, in order to probe the IR, one must also use the RG to lower the sliding scale $\Lambda$ and keep $E_i/\Lambda \sim \CO(1)$.  The additional complication here is that one may also have $\log(E_i/k_F)$ dependence,\footnote{The function $F$ can and often does also have dependence on positive powers of $\frac{k_F}{E_i}$, which also causes breakdown of perturbation theory for sufficiently small $E_i/k_F$.  However, such terms have a characteristic scale $\bar{E}$ below which they are large but far above which they are small and can be neglected. This is in contrast to $\log(E_i/k_F)$ terms, which vary equally over every change in scale. } which cannot be made small in the IR by lowering $k_F$ since $k_F$ is a physical momentum scale.  Thus, to have perturbative control over the theory over a wide range of scales, one must find a scheme where $\log(E_i/k_F)$ terms do not appear.  We may pejoratively refer to schemes that do produce $\log(E_i/k_F)$ terms as ``bad'' schemes, whereas schemes without such terms will be called ``good'' schemes.   

One example of a ``good'' scheme is to introduce a cut-off $\Lambda_k$ on the momentum of the boson in addition to the cutoff $\Lambda$ on energies.  Then, for $\Lambda_k \ll k_F$, the boson can connect only nearby fermions on the fermi surface, so as a result the $k_F$ in the fermion two-point function gets ``integrated out,'' with a one-loop result of the form
\be
{\cal A} &=& a_k \log(\Lambda_k) + a_\Lambda \log (\Lambda) + a_E \log (E),
\ee
where the $a_i$'s are functions of $\omega, \ell$,  and couplings.  Dimensional analysis implies the constraint
\be
a_k+a_\Lambda+a_E=0.
\label{eq:logmatching}
\ee
A crucial point is that different choices of regularization schemes 
can change $a_k$ and $a_\Lambda$, so that they are not unambiguous.  For instance, a different ``good'' scheme is dimensional regularization. In appendix \ref{app:OneLoop}, we briefly summarize the  calculation of the one-loop fermion Green's function in this scheme.  The result is of the form
\be
{\cal A} &=& a_\mu \log(\mu) + a_E \log(E),
\ee
where $\log (\mu)$ is the renormalization scale.  The coefficient $a_\mu$ is not equal to either $a_k$ or $a_\Lambda$ in the previous scheme.  However, what {\it is} unambiguous is the coefficient $a_E$ of the physical logarithm $\log(E)$.  The reason for this is that $\log(E)$ is a term in a physical amplitude that cannot be removed by local counter-terms and thus is unambiguously present in any regularization scheme.  In the absence of $a_k$, (\ref{eq:logmatching}) would therefore imply that $a_\Lambda = -a_E$, and so would therefore also be physical and unambiguous.  However, with $a_k$, it is possible to distribute $a_E$ among $a_k$ and $a_\Lambda$ in more than one possible way, with only their sum $a_\Lambda + a_k$ being unambiguous.

\section{Conclusions}
\label{sec:conclusions}
In this paper, we solved for the physics obtained by perturbing decoupled theories of a large $N_B$ Wilson-Fisher
boson and a Fermi liquid by a Yukawa interaction.  Using large $N_B$ techniques, we were able to provide three
concordant solutions valid at strict large $N_B$.  The RG treatment of \S2, the scaling analysis of the gap equation in
\S3, and the perturbative analysis of the gap equation in \S4 all converge to the same answer.  The fermions are
dressed into a  non-Fermi liquid where $v_F$ runs to zero, and the sub-leading term $\ell^2 / 2m^*$ in the
fermion dispersion becomes important.  In many ways, the physics is reminiscent of models of local quantum criticality
(in the regime where $v_F$ is small and the $\ell^2 / 2m^*$ is a tiny correction, as $m^*$ is very large).   However, there is a crucial difference between our large $N_B$ solution and local quantum criticality: the bosons remain at their Wilson-Fisher fixed point and therefore, boson 2-point functions are not local in the sense that they retain momentum dependence.  

At finite $N_B$, the physics we found here will break down at a low energy scale $E_{\rm breakdown}$, as discussed at
length in the introduction.  However, both because in some controlled models other instabilities may occur before
$E_{\rm breakdown}$, and because the resulting physics ${\it below}$ the scale $E_{\rm breakdown}$ will certainly
depend on the RG structure above this energy, we feel these results continue to be instructive at large but finite $N_B$.

More generally, as a function of three independent parameters -- $N_B$, $N_F$, and $\epsilon$ -- we expect that this
system admits a rich phase diagram.  In the strict $N_F \to \infty$ limit with fixed $N_B$, the results agree with those 
coming from various self-consistent ansatzes for the $\omega \to 0$ physics of this problem.  
There, Landau damping of the bosons is the dominant physics.
In the limit we took here,
which is in many ways the extreme opposite limit, the fermion self-energy is instead the most important factor.  It dresses
the fermions into a %novel 
non-Fermi liquid, and to the extent that Landau damping ever becomes important, it is very different
in form than it is in the large $N_F$ theories.  Finding controlled regions in this theory space which can be solved (other than
the strict $N_B \to \infty$ or $N_F \to \infty$ limits) is a challenging problem of great interest; the interpolation between the two extremes
is likely more physical than either one.

\acknowledgments{We acknowledge important conversations with  A. Chubukov,  S. Kivelson, and S.-S. Lee. This work was supported in part by the National Science Foundation grants PHY-0756174 (SK) and PHY-1316665 (JK), DOE Office of Basic Energy Sciences, contract DE-AC02-76SF00515 (SK and SR), a SLAC LDRD grant on `non-Fermi liquids', the John Templeton Foundation (SK and SR), and the Alfred P. Sloan Foundation (SR).  This material is based upon work supported in part by the National Science Foundation Grant No. 1066293. ALF and JK were partially supported by ERC grant BSMOXFORD no. 228169. }

\appendix

\section{Loop Effects at large $N_B$}
\label{app:Loops}

\subsection{Including $\frac{\ell^2}{2m_*}$ Corrections and the Fermion Propagator}

The fermion velocity $v$ will become very small at low energies due to the RG evolution.  In the limit that $v \to 0$ the irrelevant operator $\frac{1}{2 m_*} \psi^\dag \nabla^2 \psi$ would dominate the fermion dispersion relation, so in the small $v$ limit we must incorporate its effects.  For fermion momentum $\ell$ with
\be
v \ell \lesssim \frac{\ell^2}{2 m_*}
\ee
the new operator will become important.  Let us consider the one-loop correction to the fermion propagator in the presence of this term.  It is
\be
\CI &\equiv& g^2 \int \frac{d \omega d \ell d^2 k}{(2 \pi)^4 (\omega^2 + c^2 (\ell^2 + k^2))} \frac{1}{i(\omega+ \omega_e) - v(\ell+ \ell_e)  - \frac{(\ell + \ell_e)^2}{2 m_*}  } 
 \ee
 Now let us work in the regime of very small velocities, where $v < \frac{\ell_e}{m}$, so that the quantity
 \be
 v(\ell+ \ell_e)  + \frac{(\ell + \ell_e)^2}{2 m_*} > 0
 \ee
whenever $\ell > -\ell_e$ or $\ell < - \ell_e - \frac{1}{2} v m_*$.  We can do the $\omega$ integral by  closing the contour in the upper half $\omega$-plane.  There are two contributions, one from the boson propagator, which exists for all values of $\ell$, and one from the fermion propagator.  The contribution to $\CI$ from the boson propagator will be analytic in the complex quantity
 \be
X =  i \omega_e - v \ell_e 
 \ee
These contributions to the loop integral $\CI$ may lead to wavefunction renormalization of the fermion.  However, they cannot produce RG evolution of the velocity $v$, because the $\omega_e$ and $\ell_e$ terms are renormalized together.  

The other contribution to $\CI$ comes from the fermion propagator pole.  This contributes when $-\ell_e - \frac{1}{2} v m_* < \ell < -\ell_e$.  This pole gives
\be
 g^2 \int \frac{d^2 k}{(2 \pi)^3} \int_{-\ell_e- \frac{1}{2} v m_*}^{-\ell_e}   \frac{d \ell}{ c^2 (\ell^2 + k^2) -  \left( i \omega_e +  v(\ell+ \ell_e)  + \frac{(\ell + \ell_e)^2}{2 m_*}  \right)^2 } 
\ee
The terms proportional to $\omega_e$, $v$, and $1/m_*$ are much smaller than the terms proportional to $c$, and in particular they only contribute at order $\omega_e^2$ or $v^2$ in the small $v$ limit, so we can neglect them.  Note that this is not the same as neglecting these quantities from the beginning, since they have determined the range of integration.  The remaining integral can be easily evaluated, but it does not contribute any logarithmic UV divergences proportional to $\ell_e$ or $\omega_e$, so it does not produce any RG evolution.  We conclude that if we ever enter a regime where $v < \frac{\Lambda}{m}$, the $\beta_v$ function of $v$ vanishes and the fermion velocity remains fixed.

 \subsection{Landau Damping Including $\frac{\ell^2}{2m_*}$}
 \label{app:LDsmallv}

In order to understand Landau damping as $v\rightarrow 0$, it is crucial to include corrections from the operator $\frac{1}{2 m_*} \psi^\dag \nabla^2 \psi$.  The reason for this is that in the fermion propagator at $v=0$, it is ambiguous which side of the real axis the poles in $\omega$ fall on; in other words, it is ambiguous whether on-shell modes with a given momentum have energy above or below the fermi energy. When we include the $\frac{\ell^2}{2m_*}$ correction in the propagator, however, the dispersion relation is 
\be
\omega = \epsilon(\ell) = v \ell + \frac{\ell^2}{2m_*}.
\ee
``Holes'' in the theory occur for $\epsilon(\ell)<0$, i.e. for $-2 m_* v < \ell < 0$.  In the UV, before there is any running, we expect $m_* \sim \CO(m) \equiv \CO(\frac{k_F}{v})$, and thus the lower bound on this region at $-2m_* v \sim -\CO(k_F)$ is above the cut-off of the theory and can be neglected.  However, once $v$ runs to significantly smaller values, one then has $m_* v \ll k_F$ and indeed for some sufficiently small value of $v$ will definitely be below the cut-off of the theory.  At this point, the states that were holes, inside the fermi surface, have their energy increased above the fermi energy and can no longer be pair produced with particles.  In technical terms, poles that were below the real axis for $\omega$ get pushed to be above the real axis.  As we take $v$ smoothly to zero, therefore, Landau damping smoothly shuts off. A crucial question is what values of $v$ are sufficiently small to suppress Landau damping.  A reasonable physical expectation is that Landau damping gets suppressed when the lower-bound on momenta $-2 m_* v$ becomes comparable to the relevant energy scale in the correlator, namely the momentum $q$ of the external boson, whereas for $v \gtrsim \frac{q}{m_*}$, Landau damping should be relatively unaffected.  Let us see how this works in detail.  Keeping the $\frac{\ell^2}{2m_*}$ term in the fermion propagator, the one-loop Landau damping becomes
\be
\Pi(q_0, q) &=& g^2 \int \frac{d\omega d\ell  k_F^{d-1} d^{d-1} \hat{\Omega}}{(2\pi)^{d+1}} \frac{1}{(i \omega - v_F \ell - \frac{\ell^2}{2m_*})(i (\omega+ q_0) - v_F (\ell+q \cos \theta) - \frac{(\ell+q \cos \theta)^2}{2m_*})}. \nn\\
\ee
The integrand has two poles in $\omega$, at
\be
\omega_1 = - i \left( v_F \ell + \frac{\ell^2}{2m_*}\right) \textrm{ and } \omega_2 =-q_0  -i \left( v_F (\ell+ q\cos \theta) + \frac{(\ell+q \cos \theta)^2}{2m_*} \right).
\ee
Let us choose to close the $d \omega$ contour in the upper half-plane.  Clearly, if $v=0$, then both poles are in the lower half-plane and the integral vanishes.  However, for $v >0$, the first pole is in the upper half-plane when
\be
\ell >0 \textrm{ or } \ell<-2m_* v,
\ee
and the second pole is in the upper half-plane when
\be
\ell+ q \cos \theta > 0 \textrm{ or } \ell + q \cos \theta < -2 m_* v.
\ee
The $d\omega$ integral is non-zero when exactly only of these poles lies in the upper half-plane. If we label the contribution where the first pole is in the lower half-plane and the second pole is in the upper-half plane by $\CI_1$ and  vice versa by $\CI_2$, then we have
\be
\Pi(q_0,q) &=& \CI_1+ \CI_2, \nn\\
\CI_1 &= &\frac{g^2}{(2\pi)^d} \int_{-2 m_* v}^0 \frac{ d\ell k_F^{d-1} d^{d-1} \hat{\Omega}}{-i q_0 + v_F q \cos \theta + \frac{2 \ell q \cos \theta + q^2 \cos^2\theta}{2m_*}} \Theta(\ell+ q \cos \theta < -2 m_* v \textrm{ or } \ell+q \cos \theta > 0 ) ,\nn\\
\CI_2 &=  & \frac{g^2}{(2\pi)^d} \int_{-2 m_* v}^0  \frac{d \ell' k_F^{d-1} d^{d-1} \hat{\Omega'}}{i q_0+ v_F q \cos \theta' + \frac{2 \ell' q \cos \theta' + q^2 \cos^2\theta'}{2m_*}} \Theta(\ell'+ q \cos \theta' < -2 m_* v \textrm{ or } \ell'+q \cos \theta' > 0 ). \nn\\
\ee
In $\CI_2$, we have changed integration variables from $\ell$ to $\ell'=\ell+q \cos \theta$ and from $\theta$ to $\theta'=\pi-\theta$. Next, let us break up the integration over angles into $\cos \theta>0$ and $\cos \theta<0$.  For $\cos \theta>0$, clearly we cannot have both $\ell > -2m_* v $ and $\ell +q \cos \theta < -2 m_* v$, so
\be
\CI_1 &=& \frac{g^2}{(2\pi)^d} \int_{-{\rm min}(2 m_* v, q \cos \theta)}^0 d\ell k_F^{d-1} d^{d-1} \hat{\Omega} \frac{1}{-i q_0 + v_F q \cos \theta + \frac{2 \ell q \cos \theta + q^2 \cos^2\theta}{2m_*}}.
\ee
A similar expression hold for $\CI_2$. Now it is clear that in order for Landau damping to be suppressed by small $v$, one needs $v$ small compared to $q/m_*$.  In fact, if $v > \frac{q}{2m_*}$, then the answer is completely unaffected by the limit of integration at $-2 m_* v$ .  On the other hand, when $v \ll \frac{q}{m_*}$, we can approximate this integral by neglecting $v$ and $\frac{1}{m_*}$  in the denominator and integrating over $d \ell$ from 0 to $-2 m_* v$.  Since it is clear that there are no additional $1/v$ singularities arising from the integral in this limit, one sees that as the dimensionless ration $vm_*/q$ is taken to vanish, the integral smoothly falls to zero and Landau damping shuts off.

 \subsection{One-Loop Fermion Green's Function From Dimensional Regularization}
\label{app:OneLoop}

In this appendix, we will briefly sketch the computation of the one-loop fermion Green's function using dimensional regularization.  The integral to compute is 
\be
I &=& -\frac{g^2}{(2\pi)^{d+1}}\int \frac{d \omega d \ell d^{d-1} k}{( \omega^2 + c^2(\ell^2 + k^2))(i (\omega+\omega_e) - v(\ell+\ell_e))},
\ee
at $d=3-\epsilon$.  It is convenient to rationalize the integrand so that both terms in the denominator are positive-definite, and thus Feynman parameters may be introduced:
  \be
  I &=& \left( \frac{g^2{\rm vol}(S^{d-2})}{(2\pi)^{d+1}}\right)\frac{\pi }{2 c^{2-\epsilon} \sin (\frac{\pi \epsilon}{2})} \int \frac{ d \omega d \ell }{(\omega^2 + c^2 \ell^2)^{\frac{\epsilon}{2}} }\frac{(i (\omega+\omega_e)+v(\ell+\ell_e))}{(\omega+\omega_e^2) + v^2 (\ell+\ell_e)^2} \nn\\
&=&  
  \left( \frac{g^2{\rm vol}(S^{d-2})}{(2\pi)^{d+1}}\right) \frac{\pi }{2 c^{2-\epsilon} \sin (\frac{\pi \epsilon}{2})} \int_0^1 dx \frac{\epsilon}{2} (1-x)^{\frac{\epsilon}{2}-1} \nn\\
  && \int \frac{d \omega d \ell (i (\omega+\omega_e)+v(\ell+\ell_e))}{(\omega^2 + (v^2 x+c^2(1-x))\ell^2 + x(2 \omega \omega_e + 2 v^2 \ell \ell_e + \omega_e^2 + v^2 \ell_e^2))^{1+\frac{\epsilon}{2}}}.
  \ee
  One can integrate over $\omega$ and $\ell$ by shifting variables to complete the square in the denominator:
\be
I &=&  \left( \frac{g^2{\rm vol}(S^{d-2})}{(2\pi)^{d+1}}\right)\frac{\pi^2 }{2 c^{2-\epsilon} \sin (\frac{\pi \epsilon}{2})} \int_0^1 dx \frac{1}{\left( x v^2 + (1-x)c^2\right)^{\frac{3-\epsilon}{2}}} \frac{ \left( i \omega_e (x v^2 + (1-x)c^2)+c^3 v \ell_e\right)}{\left( x \left( \omega_e^2 (x v^2 + (1-x)c^2) + c^2 v^2 \ell_e^2 \right) \right)^{\frac{\epsilon}{2}} } \nn\\
  \ee
The expansion in small $\epsilon$ produces the log divergence term and the regularized finite piece:
\be
I &=& \frac{I_{-1}}{\epsilon} + I_0 + \CO(\epsilon).
\ee
 At leading order, we have
 \be
I_{-1} &=&\frac{g^2}{4\pi^2 c^2(c+|v|)} (i \omega_e  + \sgn(v) c \ell_e ).
\ee

The finite piece $I_0$ contains the terms that we have referred to as ``$a_E \log (E)$'' in the body of the paper. Neglecting some local terms proportional to $I_{-1}$, one obtains for $v>0$
\be
I_0 &=& - \frac{g^2}{8 \pi^2 c^2 (c^2-v^2)} \left[ 2c (v \ell_e - i \omega_e) \log \left( \frac{2 (v \ell_e - i \omega_e)}{c+v} \right) \right. \nn\\
&& \left.  - (c+v) (c \ell_e - i \omega_e) \log \left( \frac{ c \ell_e - i \omega_e}{c} \right) - (c-v) (c \ell_e + i \omega_e) \log \left( \frac{c \ell_e + i \omega_e}{c} \right) \right]  .
 \label{eq:FinalOneLoop}
\ee

\bibliography{qcmetal}

\end{document}